# Electron core ionization in compressed alkali metal cesium

V F Degtyareva

Institute of Solid State Physics Russian Academy of Sciences, Chernogolovka, Russia

E-mail: degtyar@issp.ac.ru

**Abstract.** Elements of group I in the Periodic table have valence electrons of *s*-type and are usually considered as simple metals. Crystal structures of these elements at ambient pressure are close-packed and high-symmetry of *bcc* and *fcc* types, defined by electrostatic (Madelung) energy. Diverse structures were found under high pressure with decrease of the coordination number, packing fraction and symmetry. Formation of complex structures can be understood within the model of Fermi sphere – Brillouin zone interactions and supported by Hume-Rothery arguments. With the volume decrease there is a gain in the band structure energy accompanied by a formation of many-faced Brillouin zone polyhedrons. Under compression to less than a half of the initial volume the interatomic distances become close to or smaller than the ionic radius which should lead to the electron core ionization. At strong compression it is necessary to assume that for alkali metals the valence electron band overlaps with the upper core electrons which increases the valence electron count under compression.

## 1. Introduction

The group-I elements of the Periodic table from Li to Cs at normal conditions are related to the free electron metals and adopt the body-centered cubic (*bcc*) structure. They transform under pressure to the face-centered cubic (*fcc*) structure and at significant compression to open and complex structures (see review papers [1-3] and references therein). Cesium is the heaviest alkali metal containing one valence electron 6*s* over the core with many inner electron levels. The route of structural transformations to complex and low-symmetry structures for Cs starts at relatively low pressure: *bcc* – *fcc* – *oC*84 at 2.4 GPa and 4.2 GPa, respectively. On further compression phases *tI*4 and *oC*16 occurred at 4.3 GPa and 12 GPa. Cs is the only alkali metal that is turning back to close-packed structure at 72 GPa with the double hexagonal close-packed cell (*dhcp*). Electronic configuration of alkalis is changed under pressure from free-electron type. For dense lithium it was predicted by Neaton and Ashcroft a tendency to "a pairing of the ions" [4]. The *post-fcc* high-pressure form Li-*cI*16 (at 40-60 GPa) is similar to Na-*cI*16 (104 –117 GPa) and related to more complex structures of heavy alkalis Rb-*oC*52 and Cs-*oC*84. For complex phases in Rb and Cs was suggested electron transfer *s* - *d* while the upper empty *d*-band is moving on compression downward and overlapping with the valence *s*-band. [5]. It is assumed that at high degrees of volume compression the valence electron are localized in the interstitial sites forming a '*pseudebinary*' ionic or '*electride*' compound [6].

Physical properties as electrical resistance and superconductivity of Cs under compression change essentially, as was discussed in review paper [2]. In this paper possible causes are analyzed that contribute to the formation of the complex crystal structures in cesium under pressure and compared these with structures of lighter alkalis. The electronic cause is suggested for the crystal structure formation and changes the physical properties.

## 2. Theoretical background and method of analysis

Two main energy contributions: electrostatic energy $E_{Ewald}$ and the electron band structure term $E_{BS}$ are essential for crystal structure stability of metallic phases. The former term dominates for high-symmetry close-packed structures and the latter usually favours the formation of superlattices and distorted structures. The energy of valence electrons is decreased due to a formation of Brillouin planes with a wave vector q near the Fermi level $k_F$ and opening of the energy pseudogap on these planes if $q_{hkl} \approx 2k_F$ [7]. Within a nearly free-electron model the Fermi sphere radius is defined as $k_F = (3\pi^2 z/V)^{1/3}$, where z is the number of valence electrons per atom and V is the atomic volume. This effect, known as the Hume-Rothery mechanism (or electron concentration rule), was applied to account for the formation and stability of the intermetallic phases in binary simple metal systems like Cu-Zn, and then extended and widely used to explain the stability of complex phases in various systems, from elemental metals to intermetallics [8-11].

It should be noted that two energy terms $E_{Ewald}$ and $E_{BS}$ have different dependence on the atomic volume as $V^{-1/3}$ and $V^{-2/3}$, therefore the latter term becomes more significant under compression. This is one of the reasons for the stabilization of the complex low-symmetry structures under pressure.

The stability of high-pressure phases in cesium is analyzed using a computer program BRIZ [12] that has been recently developed to construct Brillouin zones or extended Brillouin-Jones zones (BZ) and to inscribe a Fermi sphere (FS) with the free-electron radius $k_F$. The resulting BZ polyhedron consists of numerous planes with relatively strong diffraction factor and accommodates well the FS. The volume of BZ's and Fermi spheres can be calculated within this program. The BZ filling by the electron states ($V_{FS}/V_{BZ}$) is estimated by the program, which is important for understanding of electronic properties and stability of the given phase. For a classical Hume-Rothery phase $Cu_5Zn_8$, the BZ filling by electron states is equal to 93%, and is around this number for many other phases stabilized by the Hume-Rothery mechanism.

Diffraction patterns of these phases have a group of strong reflections with their $q_{hkl}$ lying near $2k_F$ and the BZ planes corresponding to these $q_{hkl}$ form a polyhedron that is very close to the FS. The FS would intersect the BZ planes if its radius, $k_F$, is slightly larger then $q_{hkl}/2$, and the program BRIZ can visualize this intersection. The ratio $2k_F/q_{hkl}$ is an important characteristic for a phase stabilized due to a Hume-Rothery mechanism. Thus, with the BRIZ program one can obtain the qualitative picture and some quantitative characteristics on how a structure matches the criteria of the Hume-Rothery mechanism.

It is important to estimate possible number of valence electrons for the structures of Hume-Rothery type assuming the Fermi sphere touches the BZ plains filling BZ polyhedra by electron states. This approach in analysis leads to suggesting an increase of valence electron numbers for compressed alkalis.

## 3. Results and discussion

High-pressure structures of cesium are considered here to analyse the effects of the Hume-Rothery mechanism on the occurrence of structural complexity in elements. The *post-fcc* phase of Cs-III exists in the narrow pressure range 4.2 - 4.3 GPa and transforms to Cs-IV with an open-packed *tI*4 structure. On further compression Cs-V and Cs-VI are formed with *oC*16 and *dhcp* structures (structural data are presented in Table 1). Relative compression $V/V_0$ is estimated assuming ambient lattice parameter for Cs-*bcc* to be $a_0 = 6.178$ Å [13]. There is significant volume reduction $\Delta V/V$ at the phase transitions II-III and III-IV that are equal to 6.4% and 6.1%, respectively [14], and there is a further large reduction in the volume of 9.3% at the IV- V transition [15]. The volume change at the V-VI transition is not significant and the mixed-phase region of Cs-V and VI is large, spanning from ~68 to ~95 GPa [16].

**Table 1.** Structure parameters of Cs phases (from [1]). Fermi sphere radius $k_F$, the ratio of $2k_F$ to Brillouin zone vectors ($2k_F/q_{hkl}$) and the filling degree of Brillouin zones by electron states $V_{FS}/V_{BZ}$ are calculated by the program BRIZ [12].

| Phase<br>Pearson symbol | Cs-III<br>oC84 | Cs-IV<br>tI4 | Cs-V<br>oC16 | Cs-IV<br>hP4 (dhcp) |
|---|---|---|---|---|
| *Structural data* | | | | |
| Space group | $C222_1$ | $I4_1/amd$ | $Cmca$ | $P6_3/mmc$ |
| Pressure | 4.3 GPa | 8 GPa | 12 GPa | 92 GPa |
| Lattice parameters (Å) | a =9.2718<br>b =13.3013<br>c =34.2025 | a =3.349<br>c =12.487 | a = 11.205<br>b =6.262<br>c =6.595 | a = 3.011<br>c =9.710 |
| Atomic volume (Å$^3$) | 50.21 | 35.01 | 30.60 | 19.06 |
| $V/V_0$ | 0.426 | 0.297 | 0.259 | 0.162 |
| *FS - BZ data from the BRIZ program* | | | | |
| z (number of valence electrons per atom) | 0.96 | 2.5 | 4 | 4 |
| $k_F$ (Å$^{-1}$) | 0.827 | 1.283 | 1.570 | 1.838 |
| Total number of BZ planes | 24 | 16 | 32 | 20 |
| $k_F/(½q_{hkl})$<br>max<br>min | <br>1.010<br>0.981 | <br>1.066<br>0.905 | <br>1.055<br>1.010 | <br>1.040<br>0.881 |
| Filling of BZ with electron states $V_{FS}/V_{BZ}$ | 0.844 | 0.792 | 0.930 | 0.786 |

*3.1. Cs-III-oC84 structure as the Hume-Rothery phase*

The *post-fcc* phase of Li at pressures ~40 GPa was found with the *cI*16 structure that is $2 \times 2 \times 2$ superlattice of *bcc* with slight shifts of atoms resulting in new (211) planes just in contact with the FS [1,2]. The new Brillouin zone is ~90 % filled by electron states, satisfying the Hume-Rothery rules. Similar phase was found for Na and for heavy alkalis the *post-fcc* phases have more complex structures such as Rb-*oC*52 and Cs-*oC*84. These structures are closely related to *cI*16 and can be considered as superstructures of *bcc* as $2 \times 2\sqrt{2} \times 3\sqrt{2}$ and $2 \times 2\sqrt{2} \times 5\sqrt{2}$, respectively. For these superstructures the number of atoms in the unit cell is N = 48 for Rb and N = 80 for Cs. However, the real number of atoms was found to be 52 and 84, respectively. Additional number of atoms can be explained by a reduction of *s* electrons due to *s* to *d* transfer. For Cs-*oC*84 the number of *s* electrons is considered to be $z = 0.96$, close to the value ~80/84, as indicated for $2k_F$ position on Figure 1 (top panel). The first diffraction peak (211) for *cI*16 of Li and Na splits in Cs-*oC*84 because of the superlattice formation into several peaks, which are close to $2k_F$ position, satisfying the Hume-Rothery effects.

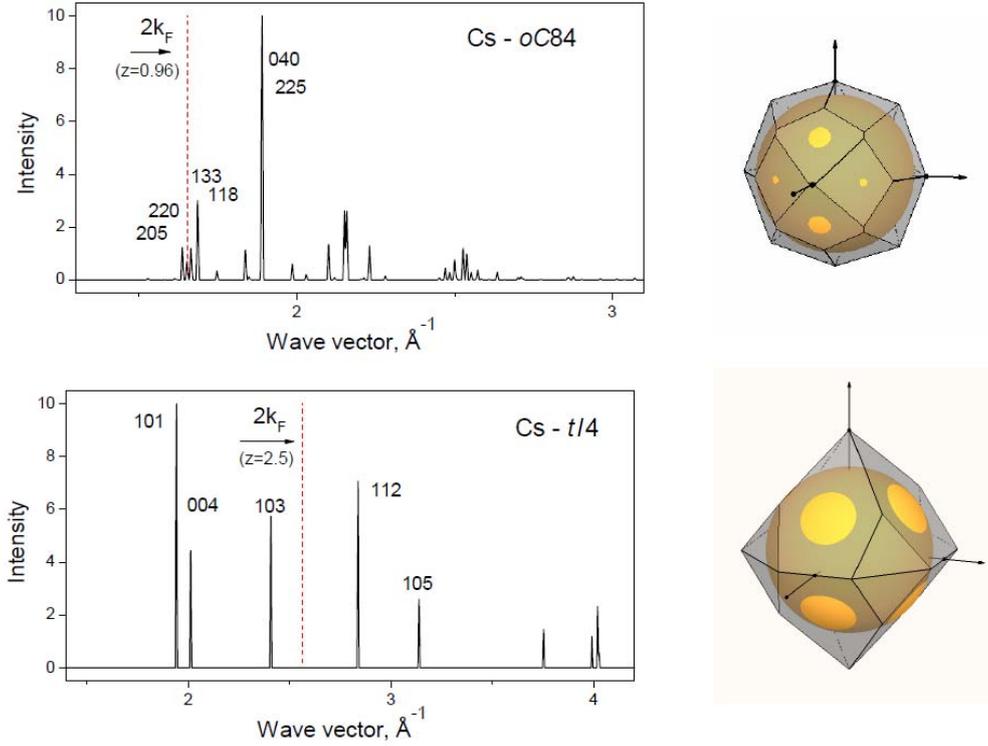

**Figure 1.** Calculated diffraction patterns (left) and corresponding Brillouin zones with the inscribed FS (right). In the top panel – for Cs-III, *oC*84 and in the bottom panel – for Cs-IV, *tI*4 with structural parameters given in table 1. The position of $2k_F$ for given z and the *hkl* indices of the principal planes are indicated on the diffraction patterns. Brillouin zones are shown in common view with a* down and c* up.

*3.2. Valence electron count in Cs-IV- tI4*

The Cs-IV structure has been observed at pressures 4.3 -12 GPa and is defined as a tetragonal cell containing 4 atoms with the space group *I*4$_1$/*amd* [13] (Pearson notation *tI*4, coordination number CN=8). That was the first case of finding this type of structure in any element and it was found also in lighter alkalis Rb and K. The volume reduction at III − IV transition is ~6% and the volume compression $V/V_0$ is about 0.3. For Cs-*tI*4 at 8 GPa interatomic distances to 4 + 4 neighbours are 3.35 Å and 3.54 Å, which are close to the double ionic radius 1.74 Å × 2 = 3.48 Å, suggested by Shannon for Cs at CN = 8 [17]. At these values a core overlapping should be expected and as consequence it is necessary to assume a transfer of upper core electrons into the valence band via *spd* hybridization of 6*s* - 5*p* - 5*d* electron levels. The effective number of valence electrons should increase and for Cs-*tI*4 it is suggested that the number of valence electrons becomes $z = 2.5$, counting only the *sp* electrons. With this value of *z* there is a matching to the Hume-Rothery rule (Figure 1, bottom panel). Axial ratio of the tetragonal cell is $c/a = 3.73$ and this value is corresponding to nearly equal distances to the neighbouring atoms which results in a suitable form of the BZ polyhedron with the (103) planes being in contact with FS.

Assumption of the electron core ionisation is reasonable for explanation of the significant volume reduction at the transition from a close-packed structure to an open-packed one with quite short interatomic distances.

*3.3. Valence electron count in Cs-V-oC16 and Cs-VI-dhcp*

In the pressure range 12 – 72 GPa, the Cs-V phase has been observed with the *oC*16 structure (see Table 1 and Figure 2, top panel). Coordination number for this structure is CN = 10-11 with the interatomic distances 3.237 – 3.612 Å at 12 GPa [15]. The double ionic radius for CN = 10 suggested for Cs by Shannon [17] equals to 1.81 Å × 2 = 3.62 Å. Thus for Cs-*oC*16 there is a significant overlap for standard size ions and standard electronic configuration needs further revision. For structure stability it is necessary to assume further core ionization – an overlap of the upper core electrons with the valence band electrons. This process is suggested to have started in the Cs-V-*tI*4 phase and is assumed to continue in next phases at higher pressures. For Cs-V-*oC*16 and Cs-VI-*dhcp* valence electron count is assumed to be z = 4 which matches the Hume-Rothery criteria.

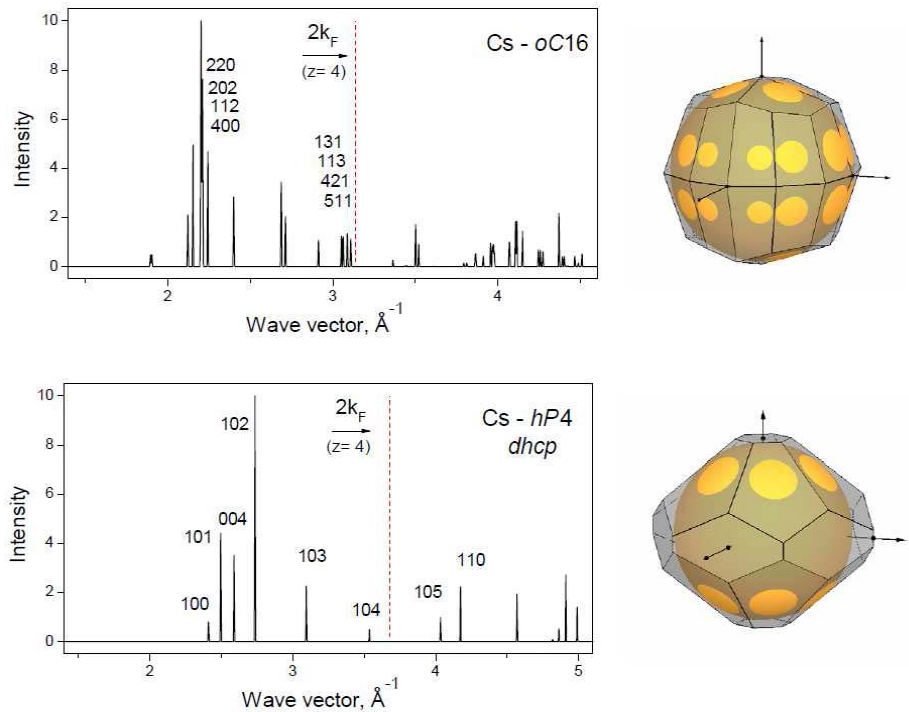

**Figure 2.** Calculated diffraction patterns (left) and corresponding Brillouin zones with the inscribed FS (right). In the top panel – for Cs-V, *oC*16 and in the bottom panel – for Cs-VI, *dhcp-hP*4 with structural parameters given in table 1. The position of $2k_F$ for given *z* and the *hkl* indices of the principal planes are indicated on the diffraction patterns.

Interestingly, that *oC*16 structure was found under pressure for group-IV elements Si and Ge that have 4 valence electrons. Similarities in atomic positions and axial ratios in the crystal structures of Cs-*oC*16, Si-*oC*16 and Ge-*oC*16 support the assumption of the similarity in electronic configurations in these elements at the pressure conditions of the *oC*16 phase, as was discussed in [9].

Next phase Cs-*VI-dhcp* also has a relation to the high-pressure phases of group-IV elements: at certain compression a close-packed hexagonal (*hcp*) structure is found in Si, Ge, Sn and Pb. The difference between *hcp* and *dhcp* is only in the different sequence of close-packed layers, it is ABA and ABACA, respectively. For Cs- *dhcp* axial ratio is *c/2a* = 1.61 that is close to the ideal *c/a* ratio for *hcp* (1.63). The reason for the occurrence of *dhcp* for Cs can be found in the stronger effects of FS-BZ interactions and by additional contributions to the band structure energy from the BZ planes (103) and (105) that are additional comparing with *hcp*.

The mixed-phase region of Cs-V and VI is quite large, spanning from ~68 to ~95 GPa which support an assumption of their nearly congruent electronic configurations and transformation is initiated by more close-packing state for *dhcp* in Cs-VI. No further phase transitions have been found and the Cs-*dhcp* has been observed up to 223 GPa, the highest pressure reached in the experimental study [18].

Physical properties for Cs under pressure correlate with the electronic state depending on the FS-BZ configuration and BZ filling by electron states, as have been analysed in the review paper [2]. For electrical resistance there is a steep increase at *fcc* to *oC*84 transition, a drop at the transition to *tI*4 and a new steep increase for *oC*16. These changes of resistivity directly follow the changes of BZ filling by electrons shown in the bottom line of Table 1. Appearance of superconductivity was found for Cs-*tI*4 and Cs-*oC*16 with the superconducting temperature $T_c$ up to 0.3 K at 12 GPa and 1.5 K at 14 GPa, respectively.

**4. Summary**

The analysis of the structural transitions in Cs, its coordination number, nearest neighbour distances, as well as the configuration of the Brillouin zone interaction with the Fermi sphere in the reciprocal space, all point to the suggestion that there is a core ionization happening in the high-pressure phases of Cs. The Cs-*tI*4 phase, it is suggested that the number of valence electrons equals to $z = 2.5$. In the higher-pressure phases *oC*16 and *dhcp*, the number of valence electrons is suggested to be $z = 4$. The heavy alkali metal Cs has a structural sequence under pressure that is completely different from the lightest alkali element Li which is connected to the differences in their electronic configurations. The complex phases of Li above 60 GPa are considered to have an increased valence electron number equal to $z = 2$ [19] whereas for Cs above 12 GPa it is suggested to be $z = 4$. This proposed change in Cs is associated with the existence of the upper empty *d*-band and inner *p*-band for heavy alkalis. Phases *tI*4 and *oC*16 were found also for K and Rb.

Special feature of Cs phase transformations is that after the transition from the close-packed into the complex and open-packed structures which are characterised by the decrease in coordination number there is re-entrance of a close-packed structure (*dhcp*) that is accompanied by the increase of the packing density. This behaviour resembles the transformation of polyvalent elements of group IV (Si, Ge, Sn) from open-packed to close-packed structures.

This analogy in structural transitions between Cs and the polyvalent elements supports the suggestion to consider similarity of valence electronic configuration for heavy alkali metals and polyvalent metals. Consideration of the core – valence band electron transfer may promote a better understanding of non-traditional behaviour of alkali elements under significant compression. Changes of physical properties of Cs under pressure could be accounted for the increase of valence electron energy contribution and, moreover, for the overlap of core and valence electron bands.

In conclusion, it might be interesting to compare the structures suggested for the metallic hydrogen at ~500 GPa [20] and Cs-V, with $I4_1/amd$- *tI*4 cells for both elements. However, in the metallic state for H is the only one valence electron and the FS radius correspond to $z = 1$. The axial ratio for H-*tI*4 is optimized as *c/a* = 2.54, whereas for Cs-V- *tI*4 *c/a* =3.73 providing the proper FS-BZ configuration with the increase of the valence electron number because of core ionization.

**Acknowledgments**

The author thanks Dr Olga Degtyareva for valuable discussions. This work was partially supported by the Program of the Russian Academy of Sciences "The Matter under High Pressure".